\documentclass[prb,aps,twocolumn]{revtex4}
\usepackage{graphicx}
\def \figurewidth{3.2in}
\begin{document}
%%%%%%%%%%%%%%%%%%%%%%%%%%%%%%%%%%%%%%%%%%%%%%%%%%%%%%%%%%%%%%%%%%%%%
 
\title{Spin Waves in Random Spin Chains}

\author{Xin Wan and Kun Yang}
\affiliation{
National High Magnetic Field Laboratory and Department of Physics,
Florida State University, Tallahassee, Florida 32306}

\author{Chenggang Zhou and R. N. Bhatt}
\affiliation{
Department of Electrical Engineering and Princeton Materials Institutes, 
Princeton University, Princeton, New Jersey 08544}

\date{\today}

\begin{abstract}

We study quantum spin-1/2 Heisenberg ferromagnetic chains with dilute,
random antiferromagnetic impurity bonds with modified spin-wave theory.
By describing thermal excitations in the language of spin waves, we
successfully observe a low-temperature Curie susceptibility due to
formation of large spin clusters first predicted by the real-space
renormalization-group approach, as well as a crossover to a pure
ferromagnetic spin chain behavior at intermediate and high temperatures.  
We compare our results of the modified spin-wave
theory to quantum Monte Carlo simulations.

\end{abstract}


\maketitle

%%%%%%%%%%%%%%%%%%%%%%%%%%%%%%%%%%%%%%%%%%%%%%%%%%%%%%%%%%%%%%%%%%%%%

One-dimensional (1D) quantum Heisenberg systems with random
ferromagnetic-antiferromagnetic (FM-AFM) couplings are of interest due
to the interplay between quantum fluctuations and disorder.  An
archetypical system is Sr$_3$CuPt$_{1-x}$Ir$_x$O$_6$,~\cite{nguyen96}
alloy of the pure compounds Sr$_3$CuPtO$_6$ (antiferromagnet) and
Sr$_3$CuIrO$_6$ (ferromagnet).  Another example is organic radical alloy
($p$-CDpOV)$_{1-x}$($p$-BDpOV)$_x$.~\cite{mukai99} These systems are
often refered to as {\it quasi}-1D random spin chains since the
interchain coupling cannot be neglected at low temperatures where
three-dimensional (3D) ordering dominates.~\cite{irons00} Curie-like
magnetic susceptibility with impurity-concentration dependent Curie
constant has been reported for these systems.~\cite{nguyen96,mukai99}

Theoretically, random spin chains have been studied by high-temperature
series expansions,~\cite{furusaki94,furusaki95} quantum transfer matrix
method,~\cite{furusaki94,furusaki95} quantum Monte Carlo (QMC)
simulations,~\cite{frischmuth97,frischmuth99,ammon99} density matrix
renormalization group method,~\cite{hida97,hikihara99} and most
noticeably, the real-space renormalization-group (RSRG)
method.~\cite{westerberg95,westerberg97} In a simple RSRG picture, the
Curie-like susceptibility occurs because ferromagnetic couplings lead to
spin correlations that form spin clusters whose average size ($\bar{N}$)
and effective spin ($\bar{S}_{eff}$) grow stochastically ($\bar{N} = A
\bar{S}_{eff}^2$) at low temperatures. The coefficient $A$ depends on
the distribution of random couplings, and, in general, increases with
concentration of ferromagnetic couplings.
Therefore, the magnetic susceptibility per spin ($\chi$) varies with 
temperature ($T$) as
\begin{equation}
\chi = { 1 \over 3 T} {\bar{S}_{eff}^2 \over \bar{N}} 
\propto {1 \over T},
\end{equation}
{\it i.e.} Curie-like, in the low-$T$ limit. Here, we assume
$\mu = k_B = 1$ for convenience. 

In the RSRG treatment, one decimates the spins that are coupled by the
strongest coupling in the system, and renormalizes the couplings among
the remaining spins perturbatively. This procedure generates effective
spins with effective (in general weaker) couplings, so that excitations
at lower and lower energy scale can be probed. It becomes obvious that
an initial broad distribution of spin couplings ensures the quick
approach to the asymptotic renormalization-group (RG) behavior. However,
it is not the case for a Heisenberg spin chain with nearest-neighboring
exchanges $+J$ or $-J$ with certain probabilities (a minimal model for
real systems).  In fact, an RSRG study of random spin
chains~\cite{westerberg97} has suggested the possibility of having a
crossover region spanning more than five orders of magnitude in $T$
before the true RSRG scaling regime can be reached. This suggests that
the crossover physics may be more relevant to experimental findings.

\begin{figure}
{\centering \includegraphics[width=\figurewidth]{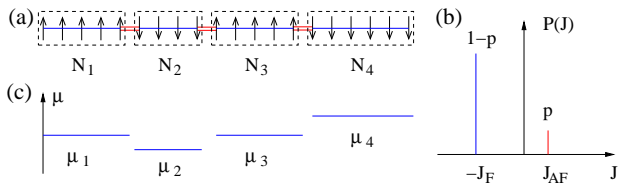} }
\caption{\label{fig1}
(a) FM segments with AFM couplings (double bonds). The segment length
$N_1$, $N_2$, ..., are exponentially distributed with an average value
$\langle N \rangle = 1/p$.
(b) Distribution of FM and AFM couplings. 
(c) In the MSW theory, local chemical potentials ($\mu_1$, $\mu_2$, ...) 
are introduced for all FM segments, determined by zero-magnetization
constraints in all segments self-consistently. }
\end{figure}

In this paper, we model a random spin chain ($S = 1/2$) by the 
following Heisenberg Hamiltonian
\begin{equation}
\label{hamiltonian}
H = \sum_i J_i {\bf S}_i \cdot {\bf S}_{i+1},
\end{equation}
where the coupling strength $J_i$ is randomly chosen between $J_{AF}$
(AFM) with probibility $p$ and $-J_F$ (FM) with probability $1-p$ (see
Fig.~\ref{fig1}).  We are primarily interested in the dilute doping
regime, {\it i.e.} $p \ll 1$, where RG-predicted Curie behavior is
expected to occur at lower $T$ with decreasing $p$. 
In the absence of AFM couplings
($J_{AF} = 0$), the system is a collection of independent FM spin segments 
with exponentially distributed numbers of spins. For small $p$, the static
magnetic susceptibility $\chi$ of the isolated spin segments shows a
crossover from an ordinary Curie behavior at high temperatures ($T >
J_F$) to a characteristic $\chi \propto 1/T^2$.  This FM-chain behavior 
has been revealed by the modified spin-wave theory.~\cite{takahashi85}
Finite amount of $J_{AF}$ introduces correlations between FM segments,
which form spin clusters that grow, in a random walk fashion, both in
size and total spin at decreasing $T$.  Statistical analysis shows that
the RG-predicted Curie susceptibility
\begin{equation}
\label{rsrg}
\chi_{RG} = {S^2 \over 3 T} {1 - p \over p}
\end{equation}
is expected at low $T$. 
Therefore, for dilute and not-too-weak AFM couplings, two crossover
between distinctive $T$-dependence can be observed in this system. 

Unfortunately, by generating effective couplings perturbatively, the
RSRG method keeps discarding high-energy excitations and, thus, is not
expected to describe correct physics well before the onset of the true
RG behavior.  To address these higher energy scales (at intermediate and
high temperatures), we have generalized the modified spin-wave theory
(MSW) to random spin systems, with dominating ferromagnetic couplings,
to study the crossover behavior from high-temperature, short-distance
physics to low-temperature, long-distance physics.  Our data suggests an
unambiguous crossover from FM spin-chain to random spin-chain behavior,
which becomes wider for smaller $p$.  Comparing with our QMC simulations
using continuous time loop cluster algorithm,~\cite{beard96} we discuss
the limitations of the MSW method.

The technical details of the application of the MSW theory in random
spin chains have been presented elsewhere.~\cite{wan02} We summarize the
MSW approach below. The main idea of the MSW theory is to introduce
a chemical potential (equivalent to a uniform magnetic field) for the
spin-wave excitations to ensure zero-magnetization of a FM spin chain at
finite temperatures, so that the number of spin waves no longer diverges
in low dimensions.~\cite{takahashi86} The introduction of the chemical
potential leads to a surprisingly good {\it quantitative} description of the
thermodynamics of the FM spin chain at both low and high $T$. In the
case of random spin chains, we introduce, as illustrated in
Fig.~\ref{fig1}(c), one such chemical potential per FM segment,
determined self-consistently by the zero-magnetization constraint of
each individual FM segment.  We apply a periodic boundary condition to
each FM segment, so that zero-magnetization of a segment implies
zero-magnetization for every spin.  With the physical significance of
restoring the rotational symmetry broken during the introduction of spin
waves, these zero-magnetization constraints ensure not only the finite
number of spin waves at finite $T$, but the appropriate correlations
between spin waves in different FM segments as well.  A bosonic version
of the generalized Bogoliubov transformation has been developed for this
approach.

\begin{figure}
{\centering \includegraphics[width=\figurewidth]{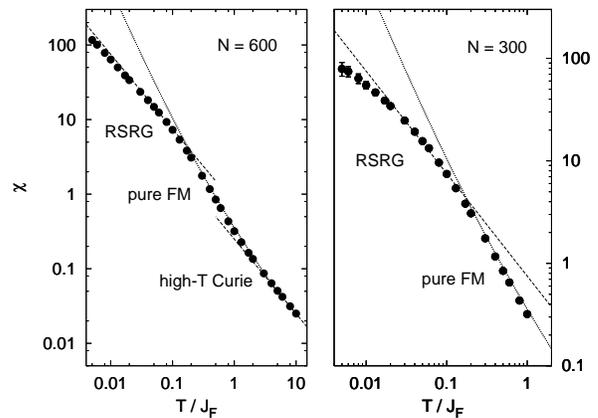} }
\caption{\label{fig2}
Left panel shows static magnetic susceptibility per spin $\chi$ of a
600-spin chain with 59 randomly distributed AFM couplings ($J_{AF} = 0.5
J_F$). On the log-log scale, the two dashed lines with slope unity are
exact results of the ordinary Curie susceptibility of independent spins
at high $T$ and the low-$T$ RG-predicted Curie susceptibility ($\chi =
0.75/T$).  The
dotted line is the low-$T$ expansion of $\chi$ of a pure spin-1/2 FM
chain (of infinite length),~\cite{takahashi85} approaching $1/T^2$ in
the low-$T$ limit (Eq.~\ref{purefm}). 
Right panel shows 7-sample averaged $\chi$ for a shorter chain of 300
spin in 30 FM segments. High-T ordinary Curie behavior is not shown. }
\end{figure}

Figure~\ref{fig2} (left panel) shows the static magnetic susceptibility
per spin $\chi$ for a random Heisenberg spin chain of 600 spins,
distributed in 30 FM segments of average length 10 spins, {\it i.e.} $p
= 0.1$, coupled by 59 AFM couplings with $J_{AF} = 0.5 J_F$.  Three
distinctive temperature regimes can be identified in the curve.  For $T
/ J_F > 3$, $\chi$ can be fit to an ordinary Curie law of independent
spins.  (We neglect this regime in later plots.)  For $0.2 < T / J_F <
3$, $\chi$ rises with decreasing $T$, following the trend of a
infinitely-long, pure FM spin-1/2 spin,~\cite{takahashi85}
\begin{equation}
\label{purefm}
\chi_{FM} = {1 \over 4} \left [ {0.1667 J_F \over T^2} 
+ {0.581 J_F^{1/2} \over T^{3/2}} + {0.68 \over T} 
+ ... \right ].
\end{equation}
This implies that spins start to correlate, forming independent
FM segments. 
The AFM couplings are still too weak to affect thermodynamics of the
random spin chain at $T > 0.2 J_F$.  
Below $T = 0.2 J_F$, $\chi$ is in good agreement with 
the low-$T$, RG-predicted Curie behavior (Eq.~\ref{rsrg} for $p = 0.1$).  
It is worth pointing out that we can identify the crossover temperature
from the FM spin-chain physics to random spin-chain physics
as the cross point of Eq.~\ref{rsrg} and \ref{purefm}. 
In the dilute doping limit (small $p$), we have 
\begin{equation}
{T_X \over J_F} = {p \over 1-p} \left [ 0.50 + 3.04 \sqrt{p \over 1-p} 
+ O\left( {p \over 1-p} \right ) \right ].
\end{equation}
For $p = 0.1$, $T_X = 0.17 J_F$.
The right panel in Fig.~\ref{fig2} shows averaged $\chi$ for 7 random
realizations of 300 spins in 30 segments with $J_{AF} = 0.5 J_F$. In
this study, we require that the total magnetization of the corresponding
classical ground state be zero, so $\chi$ vanishes at low-$T$
limit. This is responsible for the drop below $T < 0.02$, where there
are no more segments to carry out RSRG to lower $T$.  Nevertheless, the
crossover in the vicinity of $T_X = 0.17 J_F$ is clearly visible.

\begin{figure}
{\centering \includegraphics[width=\figurewidth]{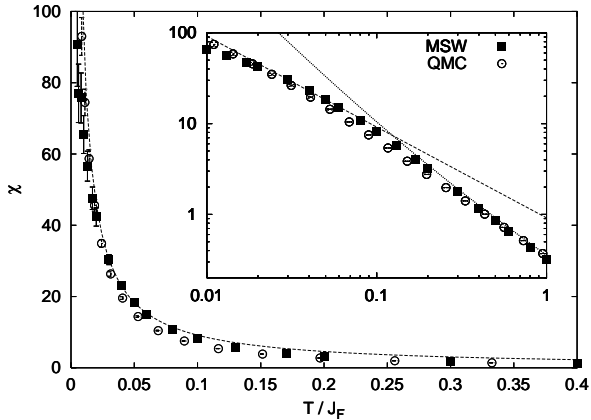} }
\caption{\label{fig3}
Avaraged susceptibility $\chi$ of random Heisenberg spin chains of 360
spins in 30 FM segments, with AFM impurity couplings $J_{AF} = 0.5 J_F$,
for the MSW method (14 samples) and the QMC simulations (28 samples). 
The dashed line ($\chi = 11/12T$) is the low-$T$ Curie susceptibility 
expected by the RSRG approach (Eq.~\ref{rsrg}).
The inset shows the same data on a log-log scale. 
The dotted line is the low-$T$ expansion of $\chi$ of a pure spin-1/2 
FM chain (of infinite length),~\cite{takahashi85} {\it i.e.} 
Eq.~\ref{purefm}. }
\end{figure}

We compare, in Fig.~\ref{fig3}, results of $\chi$ averaged over
different random realizations, obtained from both the MSW theory (14
samples) and QMC simulations (28 samples). For a system of 360 spins in
30 FM segments ($p = 1/12$) with AFM couplings ($J_{AF} = 0.5 J_F$)
between neighboring segments, we plot $\chi$ on both linear scale and
log-log scale (inset). On the log-log scale, we can easily identify the
crossover from the FM spin-chain physics to random spin-chain physics
around $T_x = 0.13 J_F$. The two methods are in good agreement,
suggesting that the MSW theory can be generalized to a certain class of
random spin chains. Near the crossover, the MSW method overestimates
$\chi$, according to the QMC simulations.  The MSW theory is
not exact even in pure FM segments; up to 5\% error can be observed for
a 12-spin FM segment near this crossover temperature.  In addition, 
by enforcing a periodic boundary condition for each FM segment, we
add an extra FM coupling per segment, which can enhance the FM behavior of
the system. We require that $\chi$ drop to zero at $T \rightarrow 0$;
the MSW method seems to feature a faster drop for such a short chain,
compared to the QMC simulations.  

The modified spin-wave theory to random spin chains has been applied to
one-dimensional Heisenberg ferrimagnets.~\cite{yamamoto98} In this case,
two constraints corresponding to zero-magnetization on each of the two
sublattices should be applied simultaneously, similar to the multiple
zero-magnetization constraints on all FM segments in random spin chains.
The linear combinations of the two chemical potentials (local fields),
in our terminology, have the physical interpretations of uniform
magnetic field and staggered magnetic field, which ensure both
magnetization and staggered magnetization be zero. The MSW study on
ferrimagnetic chains~\cite{yamamoto98} revealed that the
zero-staggered-magnetization constraint keeps number of spin waves
finite at both high-$T$ and low-$T$ limit, while the zero-magnetization
constraint does only at the low-$T$ limit. We point out that the two
chemical potentials are equivalent to the two kinds of Lagrangian
multipliers in the Schwinger-boson mean-field study of the ferrimagnetic
chains.~\cite{wu99}

This work was supported by NSF grants No. DMR-9971541 (XW and KY),
No. DMR-9809483 (CZ and RNB), the State of Florida (XW), and the
Alfred P. Sloan Foundation (KY). Three of us (XW, KY and RNB) thank
the Aspen Center for Physics for hospitality when this work has been
written up. 

%%%%%%%%%%%%%%%%%%%%%%%%%%%%%%%%%%%%%%%%%%%%%%%%%%%%%%%%%%%%%%%%%%%%%

%%%%%%%%%%%%%%%%%%%%%%%%%%%%%%%%%%%%%%%%%%%%%%%%%%%%%%%%%%%%%%%%%%%%%
\end{document}